\documentclass[prb,twocolumn,superscriptaddress]{revtex4-2}
\usepackage{amsmath,amssymb,mathrsfs}
\usepackage{graphicx}
\usepackage{colortbl}
\usepackage{braket}
\usepackage{CJK}
\usepackage{bm}
\usepackage{amsfonts}
\usepackage{gensymb}
\usepackage{xcolor}
\usepackage[normalem]{ulem}

\usepackage[colorlinks,plainpages=false,linkcolor=blue,urlcolor=blue,citecolor=blue,pdfpagemode=UseNone,pdfstartview=FitBH]{hyperref}

\begin{document}

\title{Reducing the Computational Cost Scaling of Tensor Network Algorithms via Field-Programmable Gate Array Parallelism}

\author{Songtai Lv} 
\affiliation{Key Laboratory of Polar Materials and Devices (MOE), School of Physics, East China Normal University, Shanghai 200241, China}

\author{Yang Liang} 
\affiliation{Quantum Medical Sensing Laboratory and School of Health Science and Engineering, University of Shanghai for Science and Technology, Shanghai 200093, China}

\author{Rui Zhu} 
\affiliation{Electron Microscopy Laboratory, School of Physics, Peking University, Beijing 100871, China}
\affiliation{Hefei National Laboratory, Hefei 230088, China}

\author{Qibin Zheng} 
\altaffiliation{qbzheng@usst.edu.cn}
\affiliation{Quantum Medical Sensing Laboratory and School of Health Science and Engineering, University of Shanghai for Science and Technology, Shanghai 200093, China}

\author{Haiyuan Zou} 
\altaffiliation{hyzou@phy.ecnu.edu.cn}
\affiliation{Key Laboratory of Polar Materials and Devices (MOE), School of Physics, East China Normal University, Shanghai 200241, China}

\begin{abstract}

Improving the computational efficiency of quantum many-body calculations from a hardware perspective remains a critical challenge.
Although field-programmable gate arrays (FPGAs) have recently been exploited to improve the computational scaling of algorithms such as Monte Carlo methods, their application to tensor network algorithms is still at an early stage.
In this work, we propose a fine-grained parallel tensor network design based on FPGAs to substantially enhance the computational efficiency of two representative tensor network algorithms: the infinite time-evolving block decimation (iTEBD) and the higher-order tensor renormalization group (HOTRG).
By employing a quad-tile partitioning strategy to decompose tensor elements and map them onto hardware circuits, our approach effectively translates algorithmic computational complexity into scalable hardware resource utilization, enabling an extremely high degree of parallelism on FPGAs. Compared with conventional CPU-based implementations, our scheme exhibits superior scalability in computation time, reducing the bond-dimension scaling of the computational cost from $O(D_b^3)$ to $O(D_b)$ for iTEBD and from $O(D_b^6)$ to $O(D_b^2)$ for HOTRG.
This work provides a theoretical foundation for future hardware implementations of large-scale tensor network computations.

\end{abstract}

\maketitle

\section{Introduction}

Many-body computation plays a pivotal role in condensed matter and statistical physics. Among the various computational approaches, tensor networks stand out due to their ability to efficiently capture entanglement structures and to implement controllable truncation and renormalization schemes~\cite{PEPS1,iPEPS,TEBD,MERA,TERG,Evenbly2015,Ors2014,TNreview1,Xiang2023}. Unlike sampling-based Monte Carlo methods, tensor networks introduce virtual degrees of freedom to encode entanglement, which are typically quantified by the bond dimension $D_b$. In this way, the exponential wall problem is mitigated, with the computational complexity reduced to a polynomial scaling proportional to a power of the bond dimension, $D_b^n$. Historically, tensor network methods can be viewed as a tensorial interpretation and generalization of the density-matrix renormalization group (DMRG)~\cite{DMRG} for solving many-body states~\cite{Schollwck2011}. 
With the subsequent development of tensor renormalization group (TRG) methods~\cite{TRG,Xie2012}, operator-contraction techniques have opened up alternative possibilities for investigating partition functions of quantum models with complexified parameters~\cite{liu2023CPL,liu2024PRR,Liu2024CPL,Meng2025a} and lattice field theory models in higher dimensions~\cite{Liu2013prd}, rendering tensor networks a powerful tool with broad relevance to both quantum computation and high-energy physics~\cite{Banuls2020RPP,TNreview2,WuLiu2025prl}. However, as the system dimensionality increases, the growth of the exponent $n$ severely constrains practical feasibility. At the algorithmic level, methods such as anisotropic TRG method~\cite{ATRG2020} have been proposed to reduce the computational complexity, yet they still rely heavily on parallelization using graphics processing units (GPUs) and similar architectures~\cite{Sugimoto2026}. Consequently, from a hardware perspective, the development of more effective parallelization strategies remains a key challenge for enabling large-scale tensor network computations in the future.

Traditionally, the combination of central processing units (CPUs) and GPUs in general-purpose computers has been the standard approach for achieving parallel computation~\cite{Hwu4th}. However, the intrinsic characteristics of the von Neumann architecture—such as storing instructions and data in the same memory—can lead to performance bottlenecks~\cite{Backus1978, Hennessy7th} due to limited data transfer bandwidth as well as overheads associated with instruction decoding and scheduling. Beyond general-purpose architectures, substantial efforts have been devoted to the development of application-specific integrated circuits (ASICs) to further enhance computational efficiency. For instance, tensor processing units have demonstrated outstanding capabilities in accelerating large-scale computations~\cite{Morningstar2022, Ganahl2023}. Nevertheless, the limited flexibility of ASICs, together with their problem-specific design requirements, significantly increases the development cost. In contrast, field-programmable gate arrays (FPGAs), based on classical logic gates and featuring a non–von Neumann architecture, offer both high parallelism and high flexibility. As a result, FPGAs have been widely adopted in many fields involving massive data processing~\cite{Mittal2018,Zhang2018,RodrguezBorbn2020,Shang2024} and provide a fundamental framework for the development of future computing chips~\cite{Lv2023}.

\begin{figure}[t]
\includegraphics[width=0.5\textwidth]{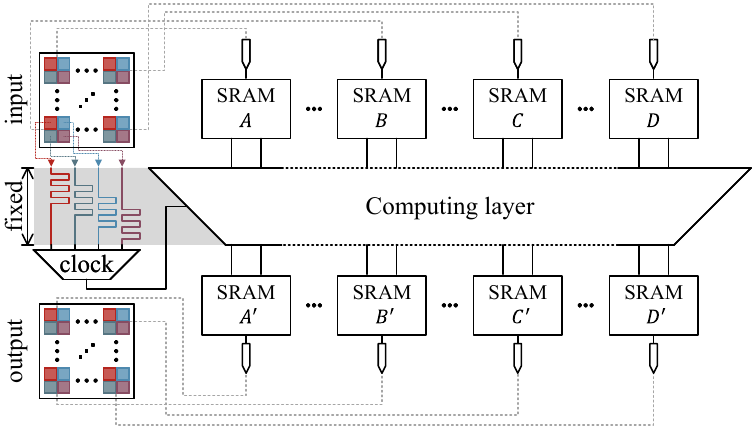}
\caption{
Schematic illustration of the parallel architecture for tensor network computations on FPGAs. The input and output tensors are partitioned into multiple small blocks, each containing a finite number of tensor elements—four elements are shown as an example and are represented by colored squares. The rectangles denote SRAMs used to store these blocks. Between the input and output, tensor elements of the same color are processed concurrently by the computing resources in the computing layer (represented by trapezoids), with a fixed number of clock cycles. All computing resources associated with each color are simultaneously driven by four pipelined time sequences. Ideally, as the number of data blocks increases, the corresponding memory and computing resources can be proportionally expanded (indicated by dashed horizontal edges of the trapezoids), while the number of clock cycles required for processing remains unchanged (indicated by solid slanted edges of the trapezoids).
}
\label{Fig.tensorToResource}
\end{figure}

Despite these advantages, the FPGA implementation of large-scale computations with many degrees of freedom, such as many-body calculations, remains challenging under current hardware resource constraints. Even so, existing studies have already demonstrated the great potential of FPGAs for specific problems~\cite{Gothandaraman08,Cardamone19,BaityJesi2013,spinglass2020,Chowdhury2023, Niazi24, Singh24}. For example, classical simulations of quantum computation based on probabilistic-bits have shown significant performance improvements over CPUs~\cite{Chowdhury2023, Niazi24, Singh24}. In Monte Carlo simulations for dense-graph problems and spin models with beyond–nearest-neighbor interactions, an $O(1)$ scaling of the sweep time—independent of the overall system size—has been achieved on FPGAs~\cite{Nikhar24,Lv2025}. Taking Monte Carlo simulations of spin models as an example, lattice degrees of freedom can be partitioned into smaller data blocks following a supercell structure~\cite{Lv2025}, and both the data blocks and their associated operations can be concurrently embedded into static random access memory (SRAM) resources such as look-up tables (LUTs) and flip-flops (FFs). Conceptually, we distinguish the data blocks and the operations by distributed SRAMs and a computing layer, respectively, as illustrated in Fig.~\ref{Fig.tensorToResource}, which presents the overall computational workflow.

For tensor network calculations, physical degrees of freedom are connected by additional virtual degrees of freedom that encode quantum entanglement, and the evaluation of physical observables requires more complex operations such as tensor contractions. The inherent nonlocality of tensor contraction—for example, matrix multiplication involving both multiplication and summation—implies that achieving an $O(1)$ scaling with respect to the bond dimension per step of tensor network evolution is theoretically impossible. Nevertheless, the framework illustrated in Fig.~\ref{Fig.tensorToResource} can be naturally extended to tensor elements. By adopting a similar block-partitioning strategy, tensor elements can still be processed in parallel, thereby maximizing computational acceleration.

Here, we take two representative tensor network algorithms—the infinite time-evolving block decimation (iTEBD) algorithm \cite{TEBD} and the higher-order tensor renormalization group (HOTRG) algorithm \cite{Xie2012}—as concrete examples to demonstrate FPGA-based implementations. These two algorithms represent two fundamental strategies in tensor network methods: imaginary-time evolution toward target states and tensor renormalization–group contractions of operators, respectively. By mapping tensor networks onto hardware circuits, we propose efficient design schemes that deeply parallelize the tensor computation structures of both algorithms on FPGAs. As a result, we achieve an $O(D_b)$ scaling of the computation time for iTEBD and an $O(D_b^2)$ scaling for HOTRG, while also surpassing GPUs in absolute performance. This work provides a novel and generally applicable parallel optimization paradigm for large-scale tensor network computations.

The rest of this paper is organized as follows. Section~\ref{sec.Methods} introduces the methodology for high-parallel tensor network computation on FPGAs based on quad-tile partitioning. Section~\ref{sec.Results} presents the computational efficiency and the corresponding hardware resource utilization, including comparisons with CPU- and GPU-based implementations. Finally, Sec.~\ref{sec.Conclusion} provides discussions and conclusions.

\section{High-parallel design} 
\label {sec.Methods}

Since iTEBD and HOTRG are two widely used tensor network algorithms, we only provide a brief overview here, and refer to Refs.~\cite{TEBD,Xie2012} for detailed descriptions. We instead focus on the key aspects relevant to parallelization, and briefly describe the hardware platform used in this work.

\subsection{Algorithm framework}

In the iTEBD algorithm, the ground-state wave function is constructed as 
\begin{equation}
\ket{\Psi}= \cdots\lambda_{2,ij}A^\alpha_{jk}\lambda_{1,kl}B^\beta_{lm}\lambda_{2,mp}\cdots\ket{\cdots\alpha\beta\cdots},
\end{equation}
where $\alpha$ and $\beta$ denote the physical indices, while $i$,$j$, etc. are virtual indices. $A$, $B$, and $\lambda$ are local tensors. The virtual indices are contracted when evaluating physical observables. The $\lambda$, serving as a simple approximation to the environment, is typically obtained through singular value decomposition (SVD). The converged ground state is then reached via iterative imaginary time evolution $U=\exp(-\tau H)$, where $H$ is the Hamiltonian of the system and $\tau$ is a tiny imaginary time. In this work, we take the one-dimensional antiferromagnetic (AF) Heisenberg model as an illustrative example. Results for other models can be readily obtained by modifying $H$, without affecting the main conclusions of this paper regarding computational efficiency.

The HOTRG algorithm is widely applied to the evaluation of the partition function, $Z = {\rm Tr}\exp(-\beta H)$ of spin systems and lattice field theories, where $H$ is the Hamiltonian and $\beta$ is the inverse temperature. The partition function can be expressed as the contraction over all indices $i$, $j$, etc, of local tensor $T_{\dots,i,j\dots}$:
\begin{equation}
Z={\rm Tr}T^N,
\end{equation}
where $N$ denotes the number of local tensors in the system.
During the contraction procedure, tensor bonds along the same direction are coarse-grained to obtain approximate tensors with reduced bond dimensions. This tensor truncation is achieved through higher-order singular value decomposition (HOSVD). In practical implementations, the tensor can be reshaped into a large matrix, such that the HOSVD step is effectively reduced to a standard matrix SVD. In contrast to iTEBD, which progressively projects the system onto the zero-temperature state, HOTRG enables access to physical quantities at arbitrary finite temperatures. For quantum spin systems, the computation of the partition function can be mapped onto a classical system with an additional imaginary-time dimension, where tensor contractions along the imaginary-time direction can be interpreted as operator evolution. 

Although the physical processes described by the iTEBD and HOTRG algorithms are different, their concrete implementations both involve two essential computational steps: tensor contraction and SVD. Parallelizing these two steps can therefore lead to a substantial acceleration of the computational efficiency of both algorithms.

\subsection{Quad-tile parallelism for tensor contraction and SVD}

In our previous works~\cite{Lv2025,Liang2025RSI}, we implemented the iTEBD algorithm on FPGAs. To the best of our knowledge, this represented the first attempt to perform tensor network calculations on an FPGA. By introducing pipelining, the resulting performance surpassed that of CPUs; however, without pipelining, the FPGA implementation was even slower than the CPU counterpart. Moreover, in both cases, the computation time still exhibited an $O(D_b^3)$ scaling with respect to the bond dimension $D_b$. In this section, we introduce a new quad-tile parallelism scheme to further optimize the two most critical computational steps: tensor contraction and SVD.

We illustrate the tensor contraction procedure using a representative example: contracting a rank-2 tensor $A$ with a rank-3 tensor $B$ over index $k$, followed by a reordering of indices to form the target tensor $M_{j[il]}$, where 
$i$ and $l$ are combined into a new composite index. Conventionally, after contracting over index $k$, one first obtains an intermediate tensor $M_{ijl}^\prime$, which must then undergo explicit permutation and reshaping operations to produce $M_{j[il]}$: 
\begin{equation}
\sum\nolimits_k A_{ik}B_{jlk}\rightarrow M_{ijl}^\prime\rightarrow M_{j[il]}.
\end{equation}
In an FPGA-oriented hardware design, however, operating directly at the level of tensor elements is more efficient than performing such high-level tensor structure manipulations~\cite{Lv2025}. Therefore, we directly map tensor elements to SRAM resources according to the $M_{j[il]}$ layout, avoiding explicit permute and reshape operations altogether.

\begin{figure}[t]
\includegraphics[width=0.5\textwidth]{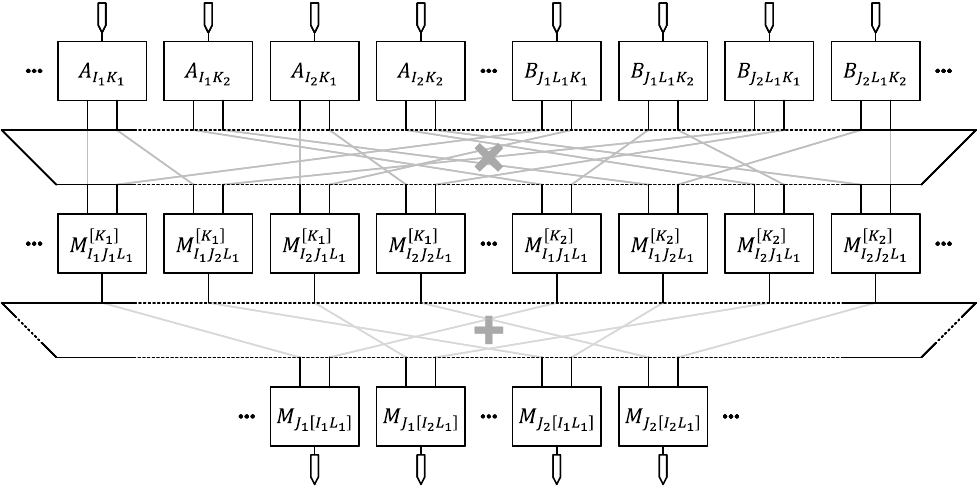}
\caption{
Schematic illustration of the FPGA logic structure and fabric configuration for tensor contraction (following the same conventions as Fig.~1 and gray lines in the compute layers indicate logical dependencies of specific input/output blocks), demonstrated using the example of multiplying two $4\times 4$ matrices according to the strategy in Eq.~(4). The input tensors $A$, $B$, and the output tensor $M$, are each partitioned into four $2\times 2$ blocks, which are individually assigned to corresponding SRAMs, e.g., $A_{I_mK_n}$. Pairwise multiplication of the block matrices produces eight intermediate blocks of size $2\times 2$. Summation of these intermediate blocks over the $K$ index then yields the output tensor $M$. For larger tensors, parallelism can be achieved by increasing the number of block indices $I_m$, $K_n$, etc. In this case, the computation time of the first computing layer remains unchanged, while the computation time of the second computing layer grows linearly with the dimension of the original tensors.
}
\label{Fig.tensorContract}
\end{figure}

Furthermore, we partition selected tensor indices into blocks with $i=i'\otimes I$, $j=j'\otimes J$, and $k=k'\otimes K$, but $l=L$. Taking the primed index to have dimension 2, for instance, each SRAM block, e.g., $A_{IK}$ and $B_{JLK}$, then contains four tensor elements, which we refer to as the quad-tile partitioning technique. With this partitioning, the original tensor contraction is decomposed into two steps, shown in Fig.~2, schematically written as,
\begin{equation}
\sum\nolimits_K\sum\nolimits_{k'}(A_{IK})_{i'k'}(B_{JLK})_{j'k'}\rightarrow (M_{J[IL]})_{i'j'}.
\end{equation}
In the first step, tensor-element multiplications together with a finite number of summations over $k'$ within each $K$ block are performed. In practice, this step is equivalent to the multiplication of two $2\times 2$ matrices. The second step performs the summation over the block index $K$. With this design, the first step can be executed fully in parallel for all blocks, and its execution time depends only on the fixed number of additions within each block, remaining constant. We assume that each tensor index has a dimension proportionally with $D_b$. Then, the second step, namely, the summation over $K$, has a computation time that scales linearly with $D_b$. As a result, the overall computational scaling is reduced from the original $O(D_b^3)$ to $O(D_b)$. We note that the summation over $K$ can be further optimized using a binary tree reduction scheme, which would reduce the scaling to $O(\ln D_b)$. However, this optimization requires additional hardware resources. Since such resources are currently limited, we do not adopt this strategy in the present work. Nevertheless, it may become advantageous in future implementations targeting larger bond dimensions.

\begin{figure}[t]
\includegraphics[width=0.5\textwidth]{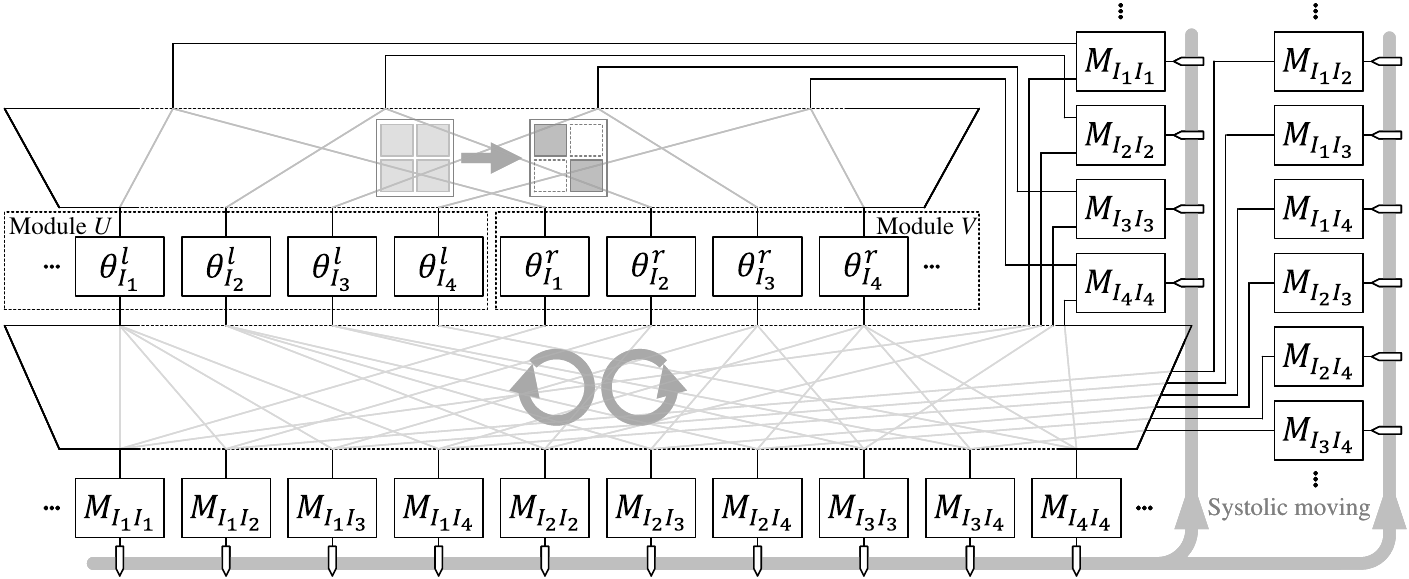}
\caption{
Schematic of the logical architecture and configuration of the SVD on FPGA (following the same conventions as Fig.~1 and gray lines in the compute layers indicate logical dependencies of specific input/output blocks). As an illustration, we consider the SVD of an $8 \times 8$ Hermitian matrix $M$. Using the quad-tile partitioning scheme, $M$ on the right is divided into four diagonal blocks $M_{I_iI_i}$ and six off-diagonal blocks $M_{I_iI_j}$ ($i\neq j$) as input tiles. $M_{I_iI_i}$ are diagonalized via Jacobi rotations (upper compute layer), producing rotation angles $\theta_{I_i}^{l/r}$, which are used to construct the modules $U$ and $V$. Applying the corresponding rotations determined by $\theta_{I_i}^{l/r}$ to all input blocks $M_{I_iI_j}$ yields the updated output blocks $M_{I_iI_j}$ (lower compute layer). After systolic data exchanges, these outputs are fed back as new inputs to iterate the above procedure. For larger matrices, parallelism can be achieved by increasing $i$ in the index $I_i$, while the computation time of both compute layers remains constant independent of $i$.
}
\label{Fig.SVD}
\end{figure}

Next, we introduce the parallel implementation of the SVD for a $D_b \times D_b$ Hermitian matrix $M$, expressed as
\begin{equation}
    M = U \Lambda V^\dagger,
\end{equation}
where $\Lambda$ is a diagonal matrix, and $U$ and $V$ are the module matrices composed of the left and right singular vectors, respectively. In our previous work~\cite{Lv2025}, although the SVD procedure was embedded into FPGA logic following a systolic array schedule~\cite{Brent1983,Luk1985}, each iteration still required operations over the entire matrix, which severely limited further improvements in computational speed.

Here, we continue to adopt the two-sided Jacobi rotation method, while simultaneously exploiting the quad-tile partitioning technique. First, $U$ and $V$ are initialized as identity matrices, and all the matrices $M$, $U$, and $V$ are partitioned into blocks. Taking $M$ as an example, each block $M_{I_i I_j}$ is a $2 \times 2$ matrix, where $I_i=1$, 2, …, $D_b/2$ labels the block indices. The left/right rotation matrix
\begin{equation}
J_{I_i}^{l/r}=
\begin{bmatrix}
\cos\theta_{I_i}^{l/r} & \sin\theta_{I_i}^{l/r}\\
-\sin\theta_{I_i}^{l/r} & \cos\theta_{I_i}^{l/r}
\end{bmatrix}
\end{equation} 
is applied to each diagonal block $M_{I_i I_i}$ such that $M’_{I_iI_i}=J_{I_i}^{l\dagger} M_{I_iI_i} J_{I_i}^r $ is diagonalized, from which the corresponding left and right rotation angles $\theta_{I_i}^l$and $\theta_{I_i}^r$ are obtained. Subsequently, the rotation matrices $J_{I_i}^{l/r}$ are applied to every block $M_{I_i I_j}$, as well as to the corresponding blocks $U_{I_i I_j}$ and $V_{I_i I_j}$. Specifically, $J_{I_i}^{l\dagger} M_{I_i I_j} J_{I_i}^r$ yields the updated matrix $M’_{I_iI_j}$, while $J_{I_i}^l U_{I_iI_j}$ and ${J_{I_i}^r V_{I_iI_j}}$ generate the updated blocks $U'_{I_i I_j}$ and $V'_{I_i I_j}$, respectively. Finally, the updated blocks $U'_{I_i I_j}$ and $V'_{I_i I_j}$ are transferred according to a systolic array schedule to construct the new module matrices $U$ and $V$.

The above procedure constitutes a single two-sided Jacobi rotation step. This step can be highly parallelized, and its execution time remains constant, independent of the matrix dimension $D_b$. After $2D_b-1$ systolic array steps, all matrix elements return to their original ordering. For a prescribed numerical precision, the above rotation steps are repeated a finite number of times until the matrix $M$ converges to a diagonal form. Consequently, the overall computational time of the SVD scales linearly with the matrix dimension $D_b$.

In the FPGA implementations of both iTEBD and HOTRG, we avoid explicit tensor reshaping and permutation operations. The execution time associated with other tasks, such as input/output data assignment, index reordering, and data-size management, is negligible compared with that of tensor contraction and SVD. Consequently, the parallel design of these two dominant operations leads to a substantial optimization of the overall efficiency of tensor network calculations.

\subsection{Hardware detail}

In principle, the above parallelization strategies can be implemented on FPGA hardware. We select the XC7K325TFFG900-2 chip manufactured by Xilinx Corporation for simulation and evaluation. The chip operates at a clock frequency of 100 MHz and is equipped with 890 block RAM (BRAM) units, 840 digital signal processing slices of the 48E version (DSP48E) slices, 407600 FFs, and 203800 LUTs. Based on this hardware configuration, simulations are performed using the Vivado High-Level Synthesis software, and the obtained results are compared with those produced by running exactly the same programs on CPU and GPU platforms. {In addition, we conduct one set of benchmark test on a real XC7K325TFFG900-2 chip.} The CPU used in our benchmarks is an Intel Xeon Gold 6230 processor with a clock frequency of 2.10 GHz, while the GPU is a Quadro K620 with a clock frequency of 1.058 GHz.

\begin{figure*}[t]
\includegraphics[width=1\textwidth]{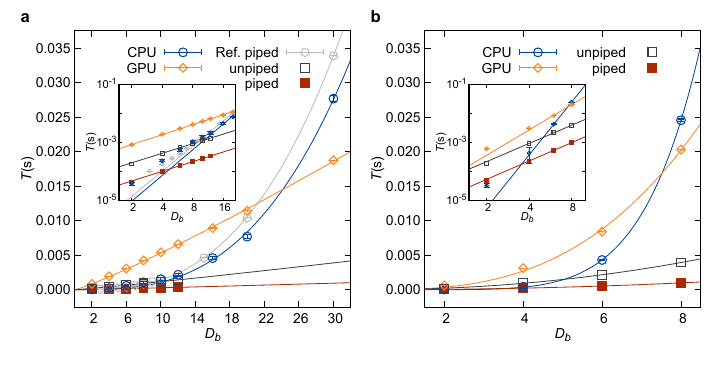}
\caption{
    The computation time {(maximum latency in simulation)} per step at different $D_b$ of (a), iTEBD calculation, and (b), HOTRG calculation for the one-dimensional AF Heisenberg chain on different platforms. The red solid squares, black hollow squares, orange diamonds, and blue circles represent the computation time for FPGA in pipelined parallel style (piped), FPGA in unpipelined parallel style (unpiped), GPU, and CPU, respectively. The error bars indicate two times the standard deviation. The solid lines represent the fitted results of the computation time, where the form of the fitting function is $D_b^{x}$ with the fitting parameter $x$. In panel (a), the fitted results for $x$ for FPGA in pipelined parallel style, FPGA in unpipelined parallel style, GPU, and CPU are 1.11, 1.09, 1.14 and 2.94, respectively. The gray pentagon denotes the computation time of FPGA in pipelined parallel style in our previous work~\cite{Lv2025} with scaling behavior $D_b^{2.88}$. In panel (b), the corresponding results of $x$ are 2.10, 2.08, 2.89 and 6.04, respectively. The insets illustrate the corresponding data and fit in log-log scale.
    }
\label{Fig.runTime}
\end{figure*}

\section{Results}
\label {sec.Results}

In this section, we compare the computational time of the iTEBD and HOTRG calculations for the one-dimensional AF Heisenberg model on CPU, GPU, and FPGA platforms. We find that the FPGA implementation achieves higher computational efficiency than both the CPU and GPU counterparts. Furthermore, we implement a pipelined parallel design on the FPGA, which provides additional speedup compared with the non-pipelined implementation. On the GPU, a multi-threaded strategy is employed to achieve parallelism, whereas the CPU calculations are carried out in a fully serial manner, such that the measured runtime directly reflects the intrinsic algorithmic complexity. In addition to the computational time, we also analyze the scaling behavior of hardware resource consumption with respect to the bond dimension $D_b$ for the two FPGA parallel implementations.

\begin{figure*}[t]
\includegraphics[width=1\textwidth]{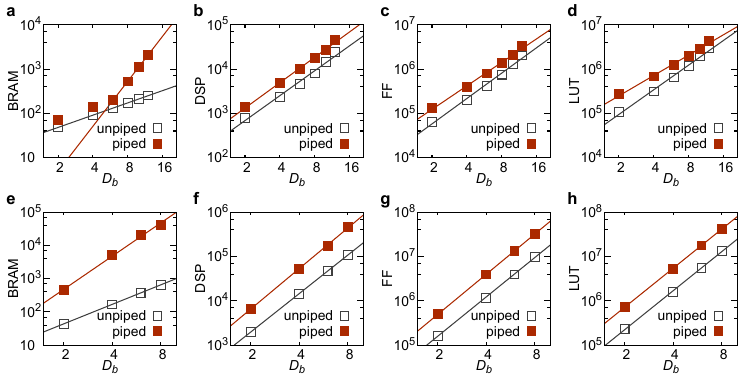}
\caption{
    The hardware resource usage at different $D_b$ of the iTEBD (a), (b), (c), (d) and the HOTRG (e), (f), (g), (h) calculation for the one-dimensional AF Heisenberg chain for two different FPGA styles. Plotting in log-log scale, the red solid squares and black hollow squares represent the hardware resource usage for FPGA in pipelined parallel style (piped) and FPGA in unpipelined parallel style (unpiped), respectively. The red and black solid lines represent the fitted results for the pipelined and unpipelined style with the fitting function $D_b^x$, respectively, where $x$ is the fitting parameter. The fitted results for $x$ of BRAM, DSP, FF and LUT for iTEBD calculation in pipelined (unpipelined) style are 3.38 (0.92), 1.91 (1.87), 1.77 (1.91) and 1.53 (1.85), respectively, and the corresponding fitted results for $x$ for HOTRG calculation are 3.32 (1.95), 3.04 (2.89), 3.01 (2.95) and 2.93 (2.93).
}
\label{Fig.resourceUsage}
\end{figure*}

In the iTEBD calculations, we simulate systems with six different bond dimensions $D_b \le 12$ on the FPGA platform. On the CPU and GPU platforms, we perform 10 000 update steps for nine different bond dimensions  $D_b \le 30$, starting from different initial states, and extract the corresponding average computation time per update step on the CPU and GPU. The results are shown in Fig.~\ref{Fig.runTime}(a). Compared with our previous FPGA computations~\cite{Lv2025}, all results exhibit substantial performance improvements. In the earlier implementation, although the pipelined FPGA version outperformed the CPU, the non-pipelined FPGA execution was slower than the CPU, and the $O(D_b^3)$ scaling behavior of the computation time with respect to $D_b$ remained unchanged. Moreover, it is worth noting that even the previously pipelined FPGA implementation was still slower than the updated CPU implementation reported here.
These observations highlight the importance of algorithm designs that are well matched to the FPGA architecture. In the earlier work, apart from pipeline parallelism, the intrinsic parallel capability of the FPGA was not fully exploited. In the present study, by employing the quad-tile partition strategy introduced in the previous section, we are able to fully leverage the parallel advantages of the FPGA. The new results demonstrate that even the non-pipelined FPGA implementation now outperforms the CPU, while the introduction of pipelining leads to a further significant speedup. More importantly, for both new FPGA implementations, the scaling of the computation time with respect to $D_b$ is dramatically reduced from $O(D_b^3)$ to $O(D_b)$, with prefactors that are also smaller than those achieved on the GPU. For instance, at $D_b = 12$, the computation speed of the FPGA with pipelined time sequencing is 19.2 times that of the GPU.
{In addition, for the iTEBD $D_b=2$ (unpiped) case where the required resources do not exceed what our real FPGA chip can provide, we conduct an actual test on the FPGA chip and obtain a time of 10789 latency per step (each latency corresponds to 10 ns), which falls between the simulated minimum latency of 5175 and maximum latency of 18 503.}

We further simulate the HOTRG calculations of the same quantum spin model on the FPGA platform. The FPGA implementation of a single update step is simulated for four different bond dimensions $D_b \le 8$, while the corresponding performance on the CPU and GPU is evaluated by averaging the execution time over 20,000 update steps. Similar to the iTEBD case, the computational efficiency on the FPGA is significantly improved compared with the CPU, and the scaling of the computation time with respect to $D_b$ is reduced from $O(D_b^6)$ on the CPU to approximately $O(D_b^2)$ on the FPGA, as illustrated in Fig.~\ref{Fig.runTime}(b). The optimization efficiency achieved for the HOTRG algorithm is conceptually consistent with that for iTEBD. The reason that the computational time scales as $D_b^2$, rather than linearly with $D_b$, originates from the intrinsic differences between the two algorithms. Unlike iTEBD, which evolves local tensor states with dimension $D_b$ along each virtual bond, the coarse-graining procedure of HOTRG along the imaginary-time direction effectively involves matrices whose row or column dimensions scale as $D_b^2$. Although the GPU implementation can also achieve a computation time scaling of $O(D_b^2)$, both the pipelined and non-pipelined FPGA implementations exhibit smaller prefactors than the GPU. For example, at $D_b=8$, the pipelined FPGA implementation achieves a computation speed that is 24.7 times that of the CPU, 20.4 times that of the GPU, and 4.0 times that of the non-pipelined FPGA implementation.

In addition to the substantial acceleration achieved by the gridding design on the FPGA, the associated hardware resource usage also exhibits favorable scaling behavior. In this work, we examine four primary types of FPGA hardware resources employed in the gridding design: BRAMs, DSP48E slices, FFs, and LUTs. As illustrated in Fig.~\ref{Fig.resourceUsage}, the utilization of all these resources follows a power-law growth for both the iTEBD and HOTRG calculations. From a functional perspective, LUTs and FFs primarily serve as on-chip SRAM resources, while DSPs are responsible for multiply–accumulate operations in tensor contractions as well as trigonometric function evaluations during the SVD procedure. In addition, LUTs, BRAMs, and FFs are jointly invoked for precomputation and data buffering. As the bond dimension $D_b$ increases, both the number of data blocks and the associated operations involved in tensor contractions and SVD scale according to power-law relations, leading to a similar scaling behavior in hardware resource utilization. Furthermore, increasing parallelism through the pipeline strategy raises overall resource consumption while preserving the power-law scaling. This behavior arises because the pipelined configuration increases the number of operations executed per unit time, thereby introducing more intermediate values and data dependencies that require additional computational and buffering resources. For the same $D_b$, HOTRG consistently consumes more hardware resources than iTEBD, since the virtual dimensions constructed and truncated in HOTRG are intrinsically larger. Moreover, BRAM usage can partially alleviate the demand for LUTs and FFs, as observed in the pipelined iTEBD implementation, where BRAM utilization increases more rapidly while the scaling of LUT and FF usage shows a slight reduction compared with the block-parallel implementation without pipelining.

\section{Conclusion and Discussion}
\label{sec.Conclusion}
In summary, we propose an effective design for highly parallel tensor network calculations on FPGAs, demonstrating substantial acceleration and favorable scalability for two representative tensor network algorithms, iTEBD and HOTRG. By mapping tensor networks onto carefully designed FPGA circuits, our approach achieves extreme parallelism, resulting in power-law reductions in computation time and performance that surpasses CPU and GPU implementations. The observed power-law scaling of hardware resource usage further underscores the feasibility of large-scale tensor network acceleration on future FPGA architectures. This work highlights the potential of FPGA-based approaches for tensor network simulations and provides a novel parallel optimization strategy for studying exotic geometric or frustrated models~\cite{LiuWY2022,Zou2023slcpl,Zou2023SCPMA}, as well as unconventional phase transitions~\cite{Wen2013,Zou2020PRB} in many-body physics.
Moreover, by establishing a direct mapping between tensor networks and hardware circuits, our study bridges tensor network algorithms and integrated circuit design, fostering mutual advances in computational physics and hardware acceleration. 
{It is worth noting that, owing to the limited resources of our current FPGA platform, the present work is primarily based on simulations rather than full FPGA hardware implementations. The actual implementations require addressing a variety of engineering issues, including routing optimization, timing closure, operating-frequency preservation, and the mitigation of bandwidth bottlenecks and communication overhead. These practical constraints ultimately determine the attainable performance on FPGA devices. A key direction of future work is to understand how algorithmic design should be adapted under such constraints. For instance, for the highly parallel Jacobi rotation scheme, it is important to determine the largest $D_b$ for which the ideal time scaling remains valid. Beyond this threshold, alternative parallelization strategies may be required, potentially leading to different scaling behaviors.
We also intend to investigate FPGA accelerators for DMRG and other tensor network algorithms, including CTMRG, and benchmark their parallel efficiency against that of conventional computing architectures~\mbox{\cite{Hager2004, Stoudenmire2013, Levy2020, Motoyama2022,Huang2021}}.}

\section*{Acknowledgments}
This work was supported by National Natural Science Foundation of China Grants (No.~12274126,~12575204) and National Key R\&D Program of China (Grant No.~2023YFF0719200).
R.Z. acknowledges the support from Innovation Program for Quantum Science and Technology (Grants No. 2021ZD0303000, and No. 2021ZD0303002).

{~\it Data availability.} The data that support the findings of this article are openly available~\cite{lv2026FPGA}.

\end{document}